# THE ROLE OF WORK-LIFE-BALANCE IN EFFECTIVE BUSINESS MANAGEMENT

Anna KASPERCZUK[1], Michał ĆWIĄKAŁA[2], Ernest GÓRKA[3], Piotr RĘCZAJSKI[4], Piotr MRZYGŁÓD[5], Maciej FRASUNKIEWICZ[6], Agnieszka DARCIŃSKA-GŁĘBOCKA[7], Jan PIWNIK[8], Grzegorz GARDOCKI[9]

[1] Bialystok University of Technology, Faculty of Mechanical Engineering; a.kasperczuk@pb.edu.pl, ORCID: 0000-0002-5919-5346
[2] I'M Brand Institute sp. z o.o.; m.cwiakala@imbrandinstitute.pl, ORCID: 0000-0001-9706-864X
[3] WSB - National-Louis University, Faculty of Social Sciences and Computer Science; ewgorka@wsb-nlu.edu.pl, ORCID: 0009-0006-3293-5670
[4] MAMASTUDIO Pawlik, Ręczajski, spółka jawna; piotr@mamastudio.pl, ORCID: 0009-0000-4745-5940
[5] Piotr Mrzygłód Sprzedaż-Marketing-Consulting; piotr@marketing-sprzedaz.pl, ORCID: 0009-0006-5269-0359
[6] F3-TFS sp. z o.o.; m.frasunkiewicz@imbrandinstitute.pl, ORCID: 0009-0006-6079-4924
[7] Bialystok University of Technology, Faculty of Mechanical Engineering; a.dardzinska@pb.edu.pl, ORCID: 0000-0002-2811-0274
[8] WSB Merito University in Gdańsk, Faculty of Computer Science and New Technologies; jpiwnik@wsb.gda.pl, ORCID: 0000-0001-9436-7142
[9] ZS "Strzelec" OSW JS 1011 Grajewo; g.gardocki@wp.pl, ORCID: 0009-0000-0220-2181
* Correspondence author

**Purpose:** This paper explores the role of work-life balance (WLB) in enhancing employee motivation. It also examines how balancing professional and personal life impacts business management.
**Design/methodology/approach**: The research used a quantitative methodology and an online survey of 102 individuals. It focused on flexible working hours, private medical care, and company cars.
**Findings:** Flexible working hours were the most effective tool for improving WLB. A positive correlation was found between WLB perception and employee motivation.
**Research limitations/implications**: The study is limited to a specific demographic and geographic scope. Future research could explore diverse cultural and occupational contexts.
**Practical implications:** Flexible working hours and private medical care significantly enhance employee satisfaction. These strategies also strengthen employer branding and reduce turnover.
**Social implications:** Promoting WLB can reduce stress and improve societal well-being. Organisations adopting WLB strategies set benchmarks for corporate social responsibility.
**Originality/value:** This study evaluates WLB tools and their impact on employee motivation. It provides valuable insights for HR managers and organisational leaders.
**Keywords:** management, work-life balance, boundary theory, motivation.
**Category of the paper:** research paper.

# 1. Introduction

In an era of rapid technological development, globalisation of markets and increasing competition, the importance of work-life balance is growing and becoming an integral part of human resource management strategies in companies around the world. Today's labour market is characterised by dynamic change, with challenges for employees to adapt, time pressures and a constant need for upskilling. In the face of these challenges, work-life balance is no longer a luxury or an add-on to employee benefits, but is becoming a key tool to influence employee motivation, productivity and loyalty (Tomaszewska-Lipiec, 2014).

Professional work, despite its potential for development and the opportunity to realise professional ambitions, often leads to excessive strain, stress and a lack of work-life balance. The negative consequences of this, such as job burnout and a decline in productivity, are increasingly recognised by employers, who realise that satisfied and motivated employees are the key to an organisation's success (Currie, Eveline, 2011).

Today, work-life balance, often referred to as work-life harmony, has become a key area of research and human resource management practices. Despite the apparent simplicity of the concept, its complexity and multidimensionality make it challenging to fully understand and effectively implement in an organisational context (Sadowska-Snarska, 2013).

The modern approach to work-life balance is based on the belief that employees should be treated not only as a source of labour, but also as individuals with a life of their own outside the workplace. Therefore, the concept promotes the harmonious integration of work and personal responsibilities, aiming to provide individuals with satisfaction and fulfilment in both the professional and private spheres (Wierda et al., 2008).

The term 'work-life balance' (WLB) gained popularity in the 1970s in the United States, mainly as a response to the challenges of the traditional family model and the increasing labour force participation of women. In the following decade, the concept expanded to include any initiative to support employees in balancing their work and personal commitments, regardless of the family context. In practice, this means that work-life harmony is achieved when neither sphere dominates at the expense of the other (Wiradendi et al., 2021).

Research shows a variety of approaches to analysing the concept of work-life balance (WLB), focusing on work-life relationships. According to the literature, there are three main approaches to describing this relationship. The first approach suggests the independence of the two spheres, according to segmentation theory. The second approach describes unidirectional influences, where mainly work interacts with private life, as presented by spillover theory. The third approach refers to reciprocal influences, according to theories such as facilitation, enrichment, domino effect and compensation theory, which implies a certain compatibility between work and personal life (Menderak, 2019).

The relationship between the spheres of work and life is often described in different ways. Firstly, they can generate conflicting expectations. Secondly, they can be mutually beneficial, where resources and experiences from one sphere benefit the other. And finally, they can be complementary, where deficiencies in one sphere are compensated by the other (Wiradendi et al., 2021).

Research on WLB in Poland often mirrors research directions observed in other parts of the world and faces similar challenges. According to the findings of Stańczak and other researchers (2017), there are six key thematic areas specific to Polish WLB research:

1. Identification of the obstacles to achieving work-life balance and the associated need for support in developing a WLB strategy.
2. Analysis of companies' approaches to WLB and their proposed solutions.
3. Exploring the experiences of women who harmonise work life with family responsibilities.
4. Defining and varying interpretations of the concept of WLB.
5. Analysing awareness of the benefits of work-life balance.
6. Evaluating the implementation of WLB strategies in practice.

## 2. Motivation theories in the context of WLB

Boundary theory emphasises the existence of two separate but related spheres: work and personal life. Depending on their values, attitudes and thinking patterns, people manage these areas variously, aiming for a harmonious combination. Balance is about minimising conflict and functioning smoothly in both areas. These boundaries can take many forms: psychological (emotions, thoughts), physical (e.g. space) or temporal (e.g. working hours) (Karassvidou, Glaveli, 2015).

Contemporary research suggests that the strength of a boundary depends on its flexibility, permeability and ability to bridge different aspects of life. Boundary crossers play an important role in shaping the work-life dynamic. Not only are they participants, but they also shape the direction and rules of both, based on their influence and identity. In the Polish scientific literature, the analysis of work-life dynamics from the perspective of boundary theory is relatively rare. Only a few publications on work-life boundaries in the context of remote working have been noted so far. There is also research on the impact of mobile technologies on work-life interactions and on work-family dynamics (Kubacka, Mroczkowska, 2020).

The research model described allows the analysis of different aspects of work-life boundaries: physical, temporal and psychological. It focuses on key variables such as interpersonal relationships or subjective interpretations and valuations in the context of organisational factors such as working time policies. In addition, it emphasises the importance of the psychological shaping and feeling of these boundaries to understand the subjective experience of balance (Kubacka, Mroczkowska, 2020).

## 3. Impact of work-life balance on commitment and job satisfaction

Maintaining a work-life balance has become an important issue in the context of quality of working life. This balance implies a smooth harmonisation of professional and personal responsibilities, which theoretically translates into improved well-being, satisfaction and efficiency at work. Being professionally active implies a deep commitment to the tasks at hand, a strong emotional bond with the organisation and a willingness to put in extra effort to achieve common goals. Research shows that employees who are able to achieve a better work-life balance tend to show greater commitment to their responsibilities. This balance also promotes better time allocation between work and private life, which reduces the risk of burnout and increases the willingness to engage with the company (Paszkiewicz, Wasiluk, 2022).

Job satisfaction is about positively evaluating one's own experiences and expectations of work. People who are able to manage their time effectively and find a work-life balance often experience higher job satisfaction. Maintaining a work-life balance can contribute to an improved sense of well-being, reduced stress levels and an increased sense of control over one's life, resulting in overall job satisfaction (Skrok et al., 2023).

For companies that care about the wellbeing of their employees, promoting work-life balance is becoming an important priority. Employees who value this balance are more loyal to their employer, less likely to change jobs and more likely to recommend it to others as an attractive place of employment. Moreover, high levels of engagement and job satisfaction translate into higher productivity, innovation and a favourable company image (Skrok et al., 2023).

## 4. Research material and method

The aim of this study was to analyse the role of work-life balance as an instrument influencing employee motivation in the context of the contemporary labour market. By exploring the mechanisms influencing work-life balance and identifying the benefits of maintaining it, an attempt was made to understand what strategies and actions can contribute to increasing employee satisfaction and commitment in the current work environment.

Research hypothesis: Organisations that actively support the work-life balance of their employees through various initiatives and programmes have more motivated employees compared to organisations that do not offer such support.

The following research questions were posed to complement the research hypothesis:
1. Are employees satisfied with their own level of work-life balance?
2. Which work-life balance instruments can be considered the most effective in the opinion of respondents?

The survey took place between May and July 2023, where the Internet was the main tool for data collection. The survey questionnaire was prepared using the Google Forms platform and made available through thematic forums and other online platforms. Participants were assured of the anonymity of their responses and the purpose of the survey was presented. In addition, participants were given the option to stop completing the questionnaire at any time.

The collected data were statistically analysed and presented in the form of tables and graphs. Frequency analysis and basic descriptive statistics for quantitative data, such as mean, median and standard deviation, were used to analyse the data. An analysis of variance (ANOVA) with repeated measures was performed to compare motivator ratings. Comparisons between two independent samples were made using the Student's t-test or the Mann-Whitney test when assumptions about the normality of the data distribution were not met. On the other hand, for comparison of values between more than two independent groups, analysis of variance (ANOVA) or the Kruskal-Wallis test was used. When significant differences were detected, POST-HOC tests were used for more detailed analysis. The normality of the data distribution was checked using the Shapiro-Wilk test. Analysis of the relationship between quantitative variables was performed using Pearson's or Spearman's correlation analysis. The significance level was taken at $\alpha = 0.05$. All analyses were performed using Statistica 13.3 software from StatSoft.

The survey involved 102 economically active people, of whom 61.76% were women and 38.24% were men. The age range of 18 to 25 years was 9.80% of the respondents, 26 to 35 years was 48.04% and 36 to 45 years was 29.41%. The remaining 12.75% were employees aged 46 and over. The largest age group in the study group was made up of respondents with a length of service of 11 to 15 years and 16 to 20 years.

## 5. Research results and discussion

An assessment of the level of work-life balance among the study participants was conducted. It was found that those surveyed generally expressed medium satisfaction with their work-life balance. This was indicated by 40.00% of respondents. However, only 6.67% of respondents were very satisfied with this balance, while the remaining 6.67% of employees admitted that they were dissatisfied with their work-life balance.

Differences in the assessment of the level of work-life balance in the light of sociodemographic variables were analysed. The results are presented in Table 1. A slightly higher assessment of work-life balance was found among women (M = 3.49; SD = 0.78) compared to men (M = 3.33; SD = 0.66). However, the demonstrated differences in the level of work-life balance between men and women did not reach statistical significance ($p > 0.05$).

**Table 1.**
*Self-assessment of work-life balance level by gender (N = 102)*

| Gender | Woman (n = 63) | | | Man (n = 39) | | | Significance |
|---|---|---|---|---|---|---|---|
| Self-assessment of Work-Life Balance | *M* | *Me* | *SD* | *M* | *Me* | *SD* | *p* |
| | 3,49 | 3,00 | 0,78 | 3,33 | 3,00 | 0,66 | 0,511 |

M – Mean; Me – Median; SD – Standard Deviation; p – Probability Level.

Source: Own elaboration based on conducted research.

In contrast, statistically significant differences were observed between age groups with regard to the level of work-life balance ($p = 0.005$). The highest level of work-life balance was observed among the oldest age group (M = 4.00; SD = 0.71). In contrast, the lowest level of self-assessed balance was observed among those aged 36-45 years (M = 3.17; SD = 0.65) (Table 2).

**Table 2.**
*Self-assessment of work-life balance level by age (N = 102)*

| Self-assessment of work-life balance level | *M* | *Me* | *SD* | *p* |
|---|---|---|---|---|
| 18–25 years (n = 10) | 3,70 | 4,00 | 0,48 | 0,005** |
| 26–35 years (n = 49) | 3,39 | 3,00 | 0,76 | |
| 36–45 years (n = 30) | 3,17 | 3,00 | 0,65 | |
| 46 years and over (n = 13) | 4,00 | 4,00 | 0,71 | |

** $p < 0.01$; M - mean; Me - median; SD - standard deviation; Z - test statistic; p - probability level.

Source: Own elaboration based on conducted research.

Respondents rated the effectiveness of selected instruments of the work-life balance concept on a scale of 1 to 5, where 1 meant 'not important', 2 - 'not very important', 3 - 'moderately important', 4 - 'quite important' and 5 - 'very important'. The results of the analysis of variance are presented in Table 3. The analysis carried out showed statistically significant differences in the level of evaluation of the instruments of the concept used in the workplace ($p < 0.001$). Instruments such as the presence of animals in the workplace (M = 1.43; SD = 1.00) and interest circles (M = 1.85; SD = 1.18) received the lowest ratings. In contrast, flexible working hours (M = 4.39; SD = 1.06) appeared to be the highest rated instrument of the concept. Additionally, the highly rated instruments were private medical care and a company car also used for personal needs.

The role of work-life-balance…                                                                 155**Table 3.**

*Evaluation of the effectiveness of selected instruments of the work-life balance concept (N = 102)*

| Instrument | M | Me | Min | Max | SD | F | p |
|---|---|---|---|---|---|---|---|
| Animals in the workplace[a] | 1,43 | 1,00 | 1,00 | 5,00 | 1,00 | 10,98 | <0,001*** |
| Hobby groups[ab] | 1,85 | 1,00 | 1,00 | 5,00 | 1,18 | | |
| Consultations with specialists on family and personal life [abc] | 2,15 | 2,00 | 1,00 | 5,00 | 1,31 | | |
| Tuition reimbursement[abcd] | 2,27 | 2,00 | 1,00 | 5,00 | 1,52 | | |
| Information on work-life balance [abcde] | 2,32 | 2,00 | 1,00 | 5,00 | 1,52 | | |
| Company nursery [abcde] | 2,37 | 2,00 | 1,00 | 5,00 | 1,56 | | |
| Study funding [bcde] | 2,59 | 2,00 | 1,00 | 5,00 | 1,70 | | |
| Mortgage assistance [bcde] | 2,69 | 3,00 | 1,00 | 5,00 | 1,43 | | |
| Family allowances [bcde] | 2,87 | 2,00 | 1,00 | 5,00 | 1,56 | | |
| Telecommuting/remote working [bcdef] | 2,95 | 3,00 | 1,00 | 5,00 | 1,60 | | |
| Transport to work [cdef] | 3,09 | 3,00 | 1,00 | 5,00 | 1,66 | | |
| Language courses [cdef] | 3,13 | 4,00 | 1,00 | 5,00 | 1,59 | | |
| Company events[cdef] | 3,17 | 3,00 | 1,00 | 5,00 | 1,40 | | |
| MultiSport card [def] | 3,33 | 3,00 | 1,00 | 5,00 | 1,49 | | |
| Catering [efg] | 3,42 | 3,00 | 1,00 | 5,00 | 1,54 | | |
| Part-time work[efg] | 3,45 | 4,00 | 1,00 | 5,00 | 1,53 | | |
| Private medical [fg] | 3,81 | 4,00 | 1,00 | 5,00 | 1,24 | | |
| Company car for personal use [fg] | 3,84 | 4,00 | 1,00 | 5,00 | 1,36 | | |
| Flexible working hours [g] | 4,39 | 5,00 | 1,00 | 5,00 | 1,06 | | |

[abc] successive letters stand for homogeneous groups; ***p < 0.001; M – mean; Me – median; SD – standard deviation; Min – minimum value; Max – maximum value; F – test statistic; p – probability level.

Source: Own elaboration based on conducted research.

The relationship between the assessment of the level of work-life balance behaviour and the perception of work motivation was analysed. The results of the analysis are presented in Table 4. A statistically significant positive correlation was found between self-assessment of the level of work-life balance and the level of work motivation (rho = 0.31; p = 0.002). The result obtained indicates that as the sense of work-life balance increases, the level of work motivation also increases.

**Table 4.**

*Relationship between work-life balance assessment and work motivation level assessment – results of correlation analysis (N = 102)*

| Work-life-balance assessment | *rho* | *p* |
|---|---|---|
| Level of motivation | 0,31** | 0,002 |

*\*\* p < 0,01; rho* – correlation coefficient; *p* – probability level.

Source: Own elaboration based on conducted research.

## 6. Summary

In today's fast-changing world, where the boundaries between work and personal life are becoming increasingly fluid, the concept of WLB is becoming extremely important for both employees and employers. Maintaining a healthy balance between these two spheres of life is crucial for an individual's wellbeing, job satisfaction and professional effectiveness. Consequently, a growing body of research is focusing on identifying the factors that influence WLB and the tools that can help improve it (Tomaszewska-Lipiec, 2014).

A study by Zheng and colleagues (2015) identified that work-life balance strategies and initiatives implemented in companies play a key role in supporting employees to achieve this balance. In addition to the programmes and initiatives offered, individual factors such as type of job, age, marital status and level of earnings influence the ability to balance work and personal life. It is therefore important to regularly monitor and adapt organisational WLB strategies and policies to better meet the individual needs of employees.

The results of the study showed that there is a slight but statistically significant difference in the assessment of WLB levels between men and women. There are several potential explanations for this phenomenon. Women may be more likely to express the need for work-life balance, which may be due to their traditional role in managing the home and caring for the family. On the other hand, men may experience more social pressure to devote more time and energy to work, which may affect their perception of their own balance (Krzykus, 2019).

The above results highlight the importance of understanding and supporting the needs of different age groups in the context of WLB. Organisations and HR decision-makers should take these differences into account when approaching policies and programmes that aim to support work-life balance.

The results of the survey, which identified flexible working hours as the highest-rated instrument in the WLB concept, reflect the growing need to adapt work schedules to the individual needs and wishes of employees. Flexible working hours enable employees to better adapt their working time to their private commitments and preferences, which can contribute to their satisfaction and commitment in the workplace. In addition, highly rated instruments such as private medical care or a company car also used for personal use emphasise the importance of employer support for aspects of life that go beyond the professional sphere (Krzykus, 2019).

The results indicate that companies are increasingly recognising and responding to employees' individual needs, offering solutions that support a harmonious work-life balance. This holistic and flexible model of approaching WLB can benefit both employees and organisations, contributing to increased productivity, employee retention and building a positive employer image.

Because of its ideological background, work-life balance (WLB) is seen as a value worthy of achievement. It is also a perceptual tool that shapes the way individuals interpret the surrounding reality, emphasising selected elements of it. The concept of WLB inspires individuals to maintain the belief that harmony can be achieved in life, and its ideological nature manifests itself in a dualistic conception of everyday life, understood as two separate domains (Ross i Vasantha, 2014).

The central idea of the WLB concept presented is to understand the ideological as well as the institutional, i.e. political, influence on the perception of work-life balance. It is these two spheres - work life and private life - that are the key areas that shape the daily choices of individuals and the relationships between them. With the concept of WLB, a clear division is made into two distinct spheres: work and private life. This classification not only makes it easier to analyse and understand individuals' daily activities, but also enables an in-depth understanding of how the two spheres interact. This division helps individuals identify areas in which they can strive to achieve harmony and balance (Skrok et al., 2023).

The introduction of the WLB concept is an important step in understanding the dynamics of work-life relationships. It provides a better understanding of how actions taken in one sphere of life can affect the other and what strategies can be adopted to make these relationships more harmonious. This concept provides a foundation for further research into work-life balance and for the development of organisational strategies and policies to support individuals in achieving this balance (Mroczkowska, Kubacka, 2020).

In conclusion, although most employees rate their work-life balance as average or satisfactory, there is a need to better understand the individual needs and expectations of employees in order to fully address the challenges of work-life harmony. Future research could also focus on the issue of work-life balance among the self-employed.

# References

1. Currie, J., Eveline, J. (2011). E-Technology and Work/Life Balance for Academics with Young Children. *Higher Education, no. 4*.
2. Karassvidou, E., Glaveli, N. (2015). Work-family balance through border theory lens: the case of a company driving in the fast lane. *Equality, Diversity and Inclusion: An International Journal, no. 1*.
3. Krzykus, J. (2019). Równowaga między pracą zawodową a życiem prywatnym – wyzwania współczesnej kobiety. In: E. Gruszewska, M. Roszkowska (eds.), *Współczesne problemy ekonomiczne w badaniach młodych naukowców. T. 3, Analizy makro- i mezoekonomiczne*.
4. Kubacka, M., Mroczkowska, D. (2020). Znaczenie technologii komunikacyjnych i mobilnych w praktykach zarządzania granicami między pracą a życiem osobistym wśród osób samozatrudnionych i pracujących na etacie. *Miscellanea Anthropologica et Sociologica, no. 1*.
5. Menderak, R. (2019). Równoważenie życia prywatnego i zawodowego pracowników w przedsiębiorstwie o strukturze macierzowej – studium przypadku. *Zeszyty Naukowe Politechniki Częstochowskiej. Zarządzanie, no. 33*.
6. Mroczkowska, D., Kubacka, M. (2020). Teorie pracy granicznej jako wyzwanie dla koncepcji work-life balance. Zarys perspektywy dla badania relacji praca–życie. *Studia Socjologiczne, no. 239*.
7. Paszkiewicz, A.J., Wasiluk, A. (2022). Motywacja do pracy osób z pokolenia Z. *Academy of Management, no. 3*.
8. Ross, D.S., Vasantha, S. (2014). A conceptual study on impact of stress on work-life balance. *Journal of Commerce & Management, no. 2*.
9. Sadowska-Snarska, C. (2013). Wspieranie równowagi praca–życie pracowników na poziomie firm: teoria i praktyka. *Prace Naukowe Uniwersytetu Ekonomicznego we Wrocławiu, no. 292*.
10. Skrok, J., Strońska-Szymanek, A., Kolemba, M., Surzykiewicz, J. (2023). Motywacja i zaspokojenie potrzeb w pracy a work-life balance i poczucie satysfakcji z życia pracowników polskich korporacji w okresie przed pandemią COVID-19. *Lubelski Rocznik Pedagogiczny, no. 2*.
11. Stańczak, A., Merecz-Kot, D., Jacukowicz, A. (2017). The comparison of Polish and Norwegian policy and research on work-life balance–current state of knowledge and future perspectives. Narrative review. *Acta Universitatis Lodziensis, Folia Sociologica, no. 61*.
12. Tomaszewska-Lipiec, R. (2014). Praca zawodowa – życie osobiste. Między harmonią a dezintegracją. In: R. Tomaszewska-Lipiec (ed.), *Relacje praca-życie pozazawodowe drogą do zrównoważonego rozwoju jednostki*. Bydgoszcz: Uniwersytet im. Kazimierza Wielkiego.
13. Wierda, B., Gerris, H., Vermulst, A.A. (2008). Adaptive Strategies, Gender Ideology, and Work Family Balance Among Dutch Dual Earners. *Journal of Marriage and Family, no. 70*.
14. Wiradendi Wolor, C., Nurkhin, A., Citriadin, Y. (2021). Is Working from Home Good for Work-Life Balance, Stress, and Productivity, or does it Cause Problems? *Humanities and Social Sciences Letters, no. 3*.
15. Zheng, C., Molineux, J., Mirshekary, M., Scarparo, S. (2015). Developing individual and organisational work-life balance strategies to improve employee health and wellbeing. *Employee Relations, no. 3*.